\documentclass{emulateapj}

\usepackage{color}
\usepackage{multirow}
\usepackage{longtable}

\usepackage{natbib}

\shorttitle{DEEP3/DEIMOS Spectroscopy of GOODS-N}
\shortauthors{Cooper et al.}

\begin{document}


\title{The DEEP3 Galaxy Redshift Survey: Keck/DEIMOS Spectroscopy in
  the GOODS-N Field\footnotemark[*]}


\author{
Michael C.\ Cooper\altaffilmark{1,\dag,2,\ddag} 
James A.\ Aird\altaffilmark{3}, 
Alison L.\ Coil\altaffilmark{3,\S}, 
Marc Davis\altaffilmark{4,5},
S.\ M.\ Faber\altaffilmark{6}, 
St\'{e}phanie Juneau\altaffilmark{2},
Jennifer M.\ Lotz\altaffilmark{7},
Kirpal Nandra\altaffilmark{8},
Jeffrey A.\ Newman\altaffilmark{9},
Christopher N.\ A.\ Willmer\altaffilmark{2},
Renbin Yan\altaffilmark{10,11}
}

\footnotetext[*]{The data presented herein
  were obtained at the W.~M.\ Keck Observatory, which is operated as a
  scientific partnership among the California Institute of Technology,
  the University of California and the National Aeronautics and Space
  Administration. The Observatory was made possible by the generous
  financial support of the W.~M.\ Keck Foundation.}

\altaffiltext{1}{Center for Galaxy Evolution, Department of Physics
  and Astronomy, University of California, Irvine, 4129 Frederick
  Reines Hall, Irvine, CA 92697, USA; m.cooper@uci.edu}

\altaffiltext{\dag}{Hubble Fellow}

\altaffiltext{2}{Steward Observatory, University of Arizona, 933 N.\
  Cherry Avenue, Tucson, AZ 85721, USA; sjuneau@as.arizona.edu,
  cnaw@as.arizona.edu}

\altaffiltext{\ddag}{Spitzer Fellow}

\altaffiltext{3}{Center for Astrophysics and Space Sciences,
  University of California, San Diego, 9500 Gilman Drive, MC 0424,
  San Diego, CA 92093, USA; acoil@ucsd.edu, jaird@ucsd.edu}

\altaffiltext{\S}{Alfred P.\ Sloan Foundation Fellow}

\altaffiltext{4}{Department of Astronomy, University of California,
  Berkeley, Mail Code 3411, Berkeley, CA 94720, USA;
  marc@astro.berkeley.edu}

\altaffiltext{5}{Department of Physics, University of California,
  Berkeley, Mail Code 7300, Berkeley, CA 94720, USA}

\altaffiltext{6}{UCO/Lick Observatory and Department of Astronomy and
  Astrophysics, University of California, Santa Cruz, 1156 High
  Street, Santa Cruz, CA 95064, USA; faber@ucolick.org,
  koo@ucolick.org, raja@ucolick.org}

\altaffiltext{7}{Space Telescope Science Institute, 3700 San Martin
  Drive, Baltimore, MD 21218, USA; lotz@stsci.edu}

\altaffiltext{8}{Max-Planck-Institut f\"{u}r Extraterrestrische Physik
  (MPE), Postfach 1312, Giessenbachstrasse 1, D-85741 Garching,
  Germany; knandra@mpe.mpg.de}

\altaffiltext{9}{Department of Physics and Astronomy, University of
  Pittsburgh, 401-C Allen Hall, 3941 O'Hara Street, Pittsbrugh, PA
  15260, USA; janewman@pitt.edu}

\altaffiltext{10}{Department of Astronomy and Astrophysics, University
  of Toronto, 50 St.\ George Street, Toronto, ON M5S 3H4, Canada}

\altaffiltext{11}{Center for Cosmology and Particle Physics,
  Department of Physics, New York University, 4 Washington Place, New
  York, NY 10003, USA; renbin@nyu.edu}

\begin{abstract}

  We present the results of spectroscopic observations in the GOODS-N
  field completed using DEIMOS on the Keck II telescope as part of the
  DEEP3 Galaxy Redshift Survey. Observations of $370$ unique targets
  down to a limiting magnitude of $R_{\rm AB} = 24.4$ yielded $156$
  secure redshifts. In addition to redshift information, we provide
  sky-subtracted one- and two-dimensional spectra of each
  target. Observations were conducted following the procedures of the
  Team Keck Redshift Survey (TKRS), thereby producing spectra that
  augment the TKRS sample while maintaining the uniformity of its
  spectral database.

\end{abstract}


\section{Introduction}
\label{sec_intro}

Due in large part to the Great Observatories Origins Deep Survey
\citep[GOODS,][]{giavalisco04}, the GOODS-N field ($\alpha = 12^{\rm
  h}36^{\rm m}55^{\rm s}$, $\delta = +62^{\circ}14^{\rm m}15^{\rm s}$)
has become one of the most well-studied extragalactic fields in the
sky with existing observations among the deepest at a broad range of
wavelengths (e.g., \citealt{alexander03}; \citealt{morrison10}; Elbaz
et al.\ in prep). In the coming years, this status as one of the
very deepest multiwavelength survey fields will be further cemented by
the ongoing and upcoming extremely-deep observations with {\it
  Spitzer}/IRAC and {\it HST}/WFC3-IR as part of the Spitzer Extended
Deep Survey (SEDS, PI G.\ Fazio) and the Cosmic Assembly Near-IR Deep
Extragalactic Legacy Survey (CANDELS, PIs S.\ Faber \& H.\ Ferguson),
respectively.

Given the large commitment of telescope time from both space- and
ground-based facilities devoted to imaging the GOODS-N field,
spectroscopic observations in this field possess a significant legacy
value. For instance, spectroscopic redshifts dramatically improve the
constraints inferred from imaging alone, allowing rest-frame
quantities to be derived with increased precision.  Furthermore, only
through spectroscopy can assorted spectral and dynamical properties
(such as the strengths and velocity widths of emission and absorption
lines) be measured.

Recognizing the potential legacy value of spectroscopic observations
in the GOODS-N field, the Team Keck Redshift Survey
\citep[TKRS,][]{wirth04} utilized the DEep Imaging Multi-Object
Spectrograph \citep[DEIMOS,][]{faber03} on the Keck II telescope to
create a publicly-available redshift catalog and uniform spectral
database across the entire area imaged with {\it HST}/ACS by the GOODS
Team. Altogether, the TKRS observed nearly $3000$ sources, yielding
secure spectroscopic redshifts for $\sim 1500$ objects and enabling
numerous studies of galaxy evolution and cosmology
\citep[e.g.,][]{kk04, weiner06, riess07, juneau10}.

In an effort to augment the value of the existing TKRS dataset, we
present observations of $370$ unique sources in the GOODS-N field (a
$> \! 10\%$ increase to the TKRS sample size), collected as part of
the DEEP3 Galaxy Redshift Survey (Cooper et al.\ 2011, in prep) and
using the same instrument and observation methods as the TKRS. The
DEEP3 survey is an ongoing spectroscopic effort designed to leverage
the vast amounts of multiwavelength data in another prime deep
extragalactic field, the Extended Groth Strip (EGS). Once completed,
DEEP3 will yield Keck/DEIMOS spectra of $\gtrsim 7500$ sources at $z <
2$, which when combined with TKRS and this work will create an
extensive spectral database that is both uniform and
publicly-available. In Sections \ref{sec_design} and \ref{sec_redux},
we describe the design, execution, and reduction of our Keck/DEIMOS
observations in GOODS-N, with the data products presented in Section
\ref{sec_data}.

\section{Target Selection and Slit Mask Design}
\label{sec_design}

The $391$ spectra presented herein were distributed across $4$
Keck/DEIMOS slit masks, designed using the DSIMULATOR software (part
of the DEIMOS IRAF package). In order to match the TKRS, targets were
selected down to a limiting magnitude of $R_{\rm AB} = 24.4$. However,
while DEIMOS $R$-band imaging was employed to select the TKRS sample,
our spectroscopic targets were drawn from the optical imaging catalogs
of \citet{capak04}.

\begin{figure}[h!]
\centering
\plotone{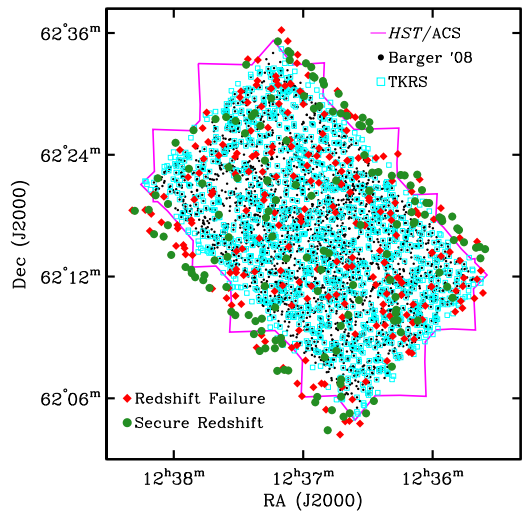}
\caption{The distribution of our Keck/DEIMOS targets on the sky. The
  cyan squares and black points show the location of sources with
  redshifts in the TKRS \citep{wirth04} and \citet{barger08} samples,
  respectively. The filled red diamonds and green circles
  denote the Keck/DEIMOS targets presented herein, with the green
  (versus red) color distinguishing those sources for which we
  succeeded (versus failed) in obtaining a secure redshift. Note that
  our target population extends beyond the GOODS {\it HST}/ACS
  footprint (denoted by the magenta outline), including previously
  unobserved {\it Spitzer} sources.}
\label{fig_skycov}
\end{figure}

Furthermore, our observations cover a moderately wider area of sky
than those of the TKRS, extending slightly beyond the borders of the
GOODS {\it HST}/ACS footprint (see Figure \ref{fig_skycov}). Like
other deep observations in GOODS-N from {\it Chandra}
\citep{alexander03} and the VLA \citep{morrison10}, the deep {\it
  Spitzer}/MIPS data from the Far-Infrared Deep Extragalactic Legacy
(FIDEL) Survey (PI M.\ Dickinson), which incorporates observations
from the {\it Spitzer} GO program of \citet{frayer06}, cover a larger
area than the {\it HST}/ACS imaging. Our slit masks were positioned,
in part, to target sources detected at $24\mu$m and/or $70\mu$m by the
FIDEL Survey, but not observed by the TKRS or as part of the
spectroscopic observations of \citet{cowie04} and
\citet{barger08}. The optical counterparts to the {\it Spitzer}/MIPS
sources were manually selected as part of the slit mask design, using
the MIPS catalogs of \citet{magnelli11}, while sources with existing
redshifts in the TKRS and \citet{barger08} catalogs were
down-weighted.

Finally, the 2 Ms {\it Chandra} data in the GOODS-N field were
reprocessed following the methodology of \citet{laird09}, with optical
counterparts to the x-ray point sources being identified in the
\citet{capak04} imaging using the likelihood ratio technique of
\citet{aird10}. Any x-ray sources lacking an existing spectroscopic
redshift in the literature, were prioritized whenever possible. The
total number of unique targets on the $4$ slit masks totals $370$,
with $21$ objects appearing on $2$ slit masks (i.e., we obtained $391$
spectra of $370$ unique targets). Note that objects targeted
specifically as {\it Spitzer} or {\it Chandra} sources are identified
accordingly in the redshift catalog (see Table \ref{tab_catalog}).

In designing the DEIMOS slit masks, slits were tilted up to $\pm 30$
degrees relative to the mask position angle to align with the major
axis of elongated targets (i.e., targets with an ellipticity $e \equiv
1 - (b/a) \ge 0.3$). When possible, an object's ellipticity and the
orientation of its major axis were estimated from the GOODS {\it
  HST}/ACS imaging using the ELLIPTICITY and THETA\_IMAGE values in
the $i_{\rm F775W}$-band imaging catalog (version r2.0z). For sources
outside of the GOODS {\it HST}/ACS footprint, the elongation and
orientation of an object were measured from the $R$-band imaging of
\citet{capak04} using the SExtractor software package
\citep{bertin96}. For non-elongated sources (i.e., $e < 0.3$), slits
were tilted $\pm 5$ degrees relative to the mask position angle so as
to provide improved wavelength sampling of the sky background.

\section{Observations and Data Reduction}
\label{sec_redux}
Spectroscopic observations using DEIMOS were completed during the
Spring of 2009 and 2010 as detailed in Table \ref{tab_obs}.
Observations utilized the 600 lines mm$^{-1}$ grating blazed at
$7500$\AA\ and tilted to a central wavelength of $7200$\AA, which
yields a nominal spectral coverage of $4600$ -- $9800$\AA\ at a
resolution (FWHM) of $\sim 3.5$\AA. The GG455 order-blocking filter
was employed to eliminate all flux blueward of $4550$\AA. Each slit
mask was observed for a total integration time of $\sim \! 3600$ sec,
divided into (at least) $3$ individual $\sim \! 1200$ sec integrations
(with no dithering performed) to facilitate the rejection of cosmic
rays --- see Table \ref{tab_obs} for details regarding the total
integration times. Standard calibration frames, consisting of three
flat-field frames utilizing the internal quartz lamp and a single arc
lamp spectrum (using Kr, Ar, Ne, and Xe), were collected for each mask
with the DEIMOS flexure compensation system ensuring that the
instrument light path for all calibration images matched the science
images to better than $\pm 0.25$ pixels.

The transparency and seeing conditions varied from poor to fair,
resuting in data of limited quality. The typical seeing varied from
$\sim 0.6^{\prime\prime}$ to $1^{\prime\prime}$ with variable cloud
cover and high humidity at times. In addition, our observations of
mask 20 were affected by intermittent dropout of the 1B CCD amplifier
in DEIMOS. This issue, which impacted two of the four exposures for
mask 20, caused all data on one-eighth of the CCD array (i.e., the
blue half of the spectrum for $\sim \! 25\%$ of all objects) to be
lost. As a result, the effective exposure time for the blue half of
the spectra in slits 0-1, 34-38, 40-41, and 44-62 of mask 20 ($28$
slits in total) is a factor of $2$ lower than that for the
corresponding red half. The resulting variation in integration time
with wavelength for each of these slits results in discontinuities in
the associated object spectra. The spectra for a fraction of these
objects still yield secure redshifts (i.e., present multiple
identifiable spectral features, with a resolved [O{\scriptsize II}]
$\lambda\lambda 3726,3729$\AA\ doublet counting as two features);
however, care must be taken in any studies utilizing information from
the blue halves of the spectra.


\begin{deluxetable*}{c c c c c c c c}
\tablewidth{0pt}
\tablecolumns{8}
\tablecaption{\label{tab_obs} Slit Mask Observation Information}
\startdata
\hline
Slit Mask & Observation Date & \multirow{2}{*}{$\alpha$ (J2000)}\footnotemark[a] &
  \multirow{2}{*}{$\delta$ (J2000)}\footnotemark[b] &
  P.A.\footnotemark[c] & \multirow{2}{*}{$N_{0}$}\footnotemark[d] & 
  \multirow{2}{*}{$N_{z}$}\footnotemark[e] & Exposure Time\footnotemark[f] \\
Number & (UT) &  &  & (deg) &  & & \\
\hline \hline
20 & 2009 Apr 22 & 12 37 48.36 & 62 06 47.86 & 45 & 118 & 50 & $4 \times 1200$s \\
21 & 2009 Apr 23 & 12 37 28.56 & 62 09 47.81 & 45 & 100 & 35 & $3 \times 1200$s \\
22 & 2009 Apr 24 & 12 36 02.88 & 62 18 51.59 & 45 & 92 & 10 & $3 \times 1200$s \\
23 & 2010 Apr 18 & 12 35 58.92 & 62 20 21.56 & 45 & 107 & 55 & $3 \times 1140$s \\
\vspace*{-0.1in}
\enddata
\tablecomments{The numbering of the DEIMOS slit masks begins at $20$
  to avoid any confusion with TKRS slit masks, which are numbered $1$
  through $18$.}

\footnotetext[a]{Right ascension (in hr mn sc) of the slit mask
  center.}

\footnotetext[b]{Declination (in deg min sec) of the slit mask
  center.}

\footnotetext[c]{Position angle of the slit mask (E of N); note that
  the orientation of individual slits vary.}

\footnotetext[d]{Number of targets on slit mask.}

\footnotetext[d]{Number of secure $(Q=-1,3,4)$ redshifts measured on
  slit mask.}

\footnotetext[d]{Total exposure time for slit mask (in seconds).}

\end{deluxetable*}

The DEIMOS spectroscopic observations were reduced using the
fully-automated DEEP2/DEIMOS data reduction pipeline (Newman et al.,
in prep; Cooper et al., in prep) developed as part of the DEEP2 Galaxy
Redshift Survey (\citealt{davis03}; Newman et al., in
prep).\footnote{http://deep.berkeley.edu/spec2d/} Redshifts were
measured from the reduced spectra using additional software developed
as part of the DEEP2 Galaxy Redshift Survey. All spectra were visually
inspected, with a quality code $(Q)$ assigned corresponding to the
accuracy of the redshift value --- $Q = -1,3,4$ denote secure
redshifts, with $Q=-1$ corresponding to stellar sources and $Q=3,4$
denoting secure galaxy redshifts (see Table \ref{tab_catalog}). For
detailed descriptions of the reduction pipeline, redshift measurement
code, and quality assignment process refer to \citet{wirth04},
\citet{davis07}, and Newman et al.\ (in prep).

\section{Data}
\label{sec_data}

\begin{figure}[h!]
\centering
\plotone{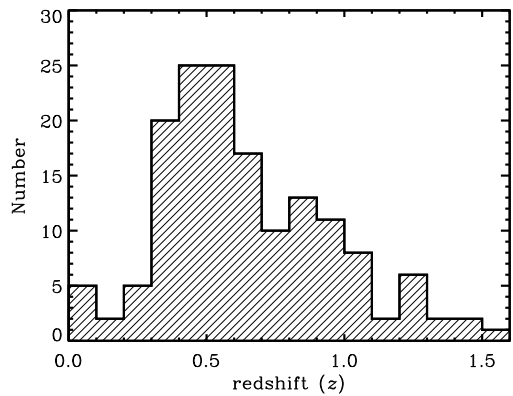}
\caption{The distribution of the $156$ unique, secure $(Q = -1,3,4)$
  redshifts measured from the Keck/DEIMOS spectroscopy.}
\label{fig_dndz}
\end{figure}

The redshift measurements resulting from our Keck/DEIMOS spectroscopy
are presented in Table \ref{tab_catalog}, a subset of which is listed
herein. The entirety of Table \ref{tab_catalog} appears in the
electronic version of the Journal and also on the DEEP Team
website.\footnote{http://deep.berkeley.edu/GOODSN} Sky-subtracted
one-dimensional and two-dimensional spectra corresponding to each
entry in the redshift catalog are also available at the same website.
Note that a redshift is only included when classified as being secure,
$(Q=-1,3,4)$. The total number of secure redshifts in the sample is
$156$ out of $370$ total, unique targets. In Figure \ref{fig_dndz}, we
show the redshift distribution for this sample. The low redshift
success rate is largely due to the poor conditions on Mauna Kea during
the observations.

Matching our catalog to those of \citet{barger08} and \citet{wirth04},
we find $34$ of our targets have a redshift published as part of these
existing data sets; $5$ of $34$ are matched to both catalogs. For $22$
of these $34$ objects, we measure a secure redshift from our DEIMOS
spectroscopy. While this sample is quite small, the agreement between
our redshifts and those of \citet{barger08} and \citet{wirth04} is
excellent. We find a median offset of $|\Delta z| \sim \! 70$ km
s$^{-1}$ and a maximum difference of $530$ km s$^{-1}$.

The new redshifts presented here should significantly enhance studies
of galaxy evolution and cosmology in the GOODS-N field. Our sample
expands upon the work of the Team Keck Redshift Survey, increasing the
size of the existing TKRS redshift and spectral data sets by
approximately $10\%$. In addition, our observations broaden the area
covered by the TKRS to extend beyond the GOODS {\it HST}/ACS
footprint, allowing us to target a greater number of relatively rare
sources.

In particular, we specifically targeted {\it Spitzer}/MIPS and {\it
  Chandra} sources not previously observed by TKRS and other
spectroscopic efforts in the field \citep[e.g.,][]{lowenthal97,
  phillips97, cohen00, dawson01, treu05, reddy06, barger08}. Within
the FIDEL Survey's {\it Spitzer}/MIPS $70\mu$m photometric catalog for
GOODS-N, there are less than 100 sources with a $5$-$\sigma$ detection
down to $3.2$ mJy \citep{magnelli11}. The relatively small number of
these sources puts a premium on spectroscopic follow-up, including
those located outside of the area imaged with {\it HST}/ACS. The
70$\mu$m observations conducted as part of the FIDEL Survey are the
deepest in the sky, allowing significant numbers of star-forming
galaxies and active galactic nuclei to be detected out to intermediate
redshift at rest-frame wavelengths that are dramatically less impacted
by aromatic and silicate emission than those normally probed by {\it
  Spitzer}/MIPS 24$\mu$m observations.  With accompanying redshift
information from spectroscopic follow-up such as presented here, these
deep far-infrared data provide a unique constaint on the cosmic
star-formation history at intermediate redshift
\citep[e.g.,][]{magnelli09}. 

Finally, by extending beyond the {\it HST}/ACS footprint (i.e., the
field surveyed by TKRS and \citealt{barger08}), this work has taken an
initial step towards expanding the area over which galaxy overdensity
(or ``environment'') can be measured in the GOODS-N field. The finite
area of sky covered by a survey introduces geometric distortions ---
or edge effects --- which bias environment measures near borders (or
holes) in the survey field, generally leading to an underestimate of
the local overdensity \citep{cooper05, cooper06}. To minimize the
impact of these edge effects on studies of galaxy environment,
galaxies near the edge of the survey field (e.g., within a projected
distance of $1$-$2\ h^{-1}$ comoving Mpc of an edge) are often
excluded from any analysis. Thus, the data presented herein, when
combined with additional spectroscopic observations that similarly
broaden the survey field, will allow the environment of galaxies at
intermediate redshift to be accurately computed across the entire {\it
  HST}/ACS area in the GOODS-N field, enabling unique studies of
small-scale clustering in one of the most well-studied extragalactic
fields in the sky.

\begin{deluxetable*}{c c c c c c c c c c c c c}
\tablewidth{0pt}
\tablecolumns{13}
\tablecaption{\label{tab_catalog} Redshift Catalog}
\tablehead{Object ID\footnotemark[a] & $\alpha$\footnotemark[b]
  (J2000) & $\delta$\footnotemark[c] (J2000) & $R_{\rm
    AB}$\footnotemark[d] & Mask\footnotemark[e] & Slit\footnotemark[f]
  & MJD\footnotemark[g] & flag\footnotemark[h] & $z$\footnotemark[i] & $z_{\rm
    helio}$\footnotemark[j] & $Q$\footnotemark[k] & $z_{\rm
    other}$\footnotemark[l] & Ref\footnotemark[m]} 
\startdata
15018 & 189.20270 & 62.073785 & 22.05 & 20 & 000 & 54943.3 &
0 & 0.2856 & 0.2855 & 4 & \ldots & \ldots \\
26812 & 189.30411 & 62.174627 & 23.27 & 20 & 007 & 54943.3 & 
0 & 0.8579 & 0.8579 & 4 & 0.8576 & 2 \\
27336 & 189.25811 & 62.155796 & 24.31 & 20 & 015 & 54943.3 & 
0 & \ldots & \ldots & 2 & \ldots & \ldots \\
\vspace*{-0.1in}
\enddata
\tablecomments{Table \ref{tab_catalog} is presented in its entirety in
  the electronic edition of the Journal. A portion is shown here for
  guidance regarding its form and content.}

\footnotetext[a]{Object identification number in $R$-band catalog of
  \citet{capak04}.}

\footnotetext[b]{Right ascension in decimal degrees from
  \citet{capak04}.}

\footnotetext[c]{Declination in decimal degrees from
  \citet{capak04}.}

\footnotetext[d]{$R$-band magnitude in AB system from
  \citet{capak04}.}

\footnotetext[e]{Number of DEIMOS slit mask on which object was
  observed.}

\footnotetext[f]{Number of slit on DEIMOS slit mask corresponding to
  object.}

\footnotetext[g]{Modified Julian Date of observation.}

\footnotetext[h]{Targeting Flag: $2 = {\rm x}$-ray target, $1 = {\rm
    MIPS}$ target, $0 = {\rm Main}$ $R$-band selected target.}

\footnotetext[i]{Redshift derived from observed spectrum.}

\footnotetext[j]{Heliocentric-frame redshift.}

\footnotetext[k]{Redshift quality code (${\rm star} = -1$; $\sim \!
  90\%\ {\rm confidence} = 3$; $\sim \! 99\%\ {\rm confidence} = 4$;
  ${\rm unknown} = 1,2$).}

\footnotetext[l]{Alternate redshift from literature.}

\footnotetext[m]{Source of alternate redshift: (1) \citet{wirth04};
  (2) \citet{barger08}.}

\end{deluxetable*}

\vspace*{0.25in} 

\acknowledgments Support for this work was provided by NASA through
the Spitzer Space Telescope Fellowship Program. MCC acknowledges
support for this work provided by NASA through Hubble Fellowship grant
\#HF-51269.01-A awarded by the Space Telescope Science Institute,
which is operated by the Association of Universities for Research in
Astronomy, Inc., for NASA, under contract NAS 5-26555. This work was
also supported in part by NSF grants AST-0808133, AST-0807630, and
AST-0806732 as well as {\it Hubble Space Telescope} Archival grant,
HST-AR-10947.01. MCC thanks Greg Wirth and the entire Keck Observatory
staff for their help in the acquisition of the DEEP3 Keck/DEIMOS data.

We also wish to recognize and acknowledge the highly significant
cultural role and reverence that the summit of Mauna Kea has always
had within the indigenous Hawaiian community. It is a privilege to be
given the opportunity to conduct observations from this mountain.

{\it Facilities:} \facility{Keck:II (DEIMOS)}



\begin{thebibliography}{28}
\expandafter\ifx\csname natexlab\endcsname\relax\def\natexlab#1{#1}\fi

\bibitem[{{Aird} {et~al.}(2010)}]{aird10}
{Aird}, J. {et~al.} 2010, \mnras, 401, 2531

\bibitem[{{Alexander} {et~al.}(2003)}]{alexander03}
{Alexander}, D.~M. {et~al.} 2003, \aj, 126, 539

\bibitem[{{Barger} {et~al.}(2008){Barger}, {Cowie}, \& {Wang}}]{barger08}
{Barger}, A.~J., {Cowie}, L.~L., \& {Wang}, W. 2008, \apj, 689, 687

\bibitem[{{Bertin} \& {Arnouts}(1996)}]{bertin96}
{Bertin}, E. \& {Arnouts}, S. 1996, \aaps, 117, 393

\bibitem[{{Capak} {et~al.}(2004)}]{capak04}
{Capak}, P. {et~al.} 2004, \aj, 127, 180

\bibitem[{{Cohen} {et~al.}(2000){Cohen}, {Hogg}, {Blandford}, {Cowie}, {Hu},
  {Songaila}, {Shopbell}, \& {Richberg}}]{cohen00}
{Cohen}, J.~G., {Hogg}, D.~W., {Blandford}, R., {Cowie}, L.~L., {Hu}, E.,
  {Songaila}, A., {Shopbell}, P., \& {Richberg}, K. 2000, \apj, 538, 29

\bibitem[{{Cooper} {et~al.}(2005){Cooper}, {Newman}, {Madgwick}, {Gerke},
  {Yan}, \& {Davis}}]{cooper05}
{Cooper}, M.~C., {Newman}, J.~A., {Madgwick}, D.~S., {Gerke}, B.~F., {Yan}, R.,
  \& {Davis}, M. 2005, \apj, 634, 833

\bibitem[{{Cooper} {et~al.}(2006)}]{cooper06}
{Cooper}, M.~C. {et~al.} 2006, \mnras, 370, 198

\bibitem[{{Cowie} {et~al.}(2004){Cowie}, {Barger}, {Hu}, {Capak}, \&
  {Songaila}}]{cowie04}
{Cowie}, L.~L., {Barger}, A.~J., {Hu}, E.~M., {Capak}, P., \& {Songaila}, A.
  2004, \aj, 127, 3137

\bibitem[{{Davis} {et~al.}(2003)}]{davis03}
{Davis}, M. {et~al.} 2003, in Society of Photo-Optical Instrumentation
  Engineers (SPIE) Conference Series, Vol. 4834, Society of Photo-Optical
  Instrumentation Engineers (SPIE) Conference Series, ed. {P.~Guhathakurta},
  161--172

\bibitem[{{Davis} {et~al.}(2007)}]{davis07}
{Davis}, M. {et~al.} 2007, \apjl, 660, L1

\bibitem[{{Dawson} {et~al.}(2001){Dawson}, {Stern}, {Bunker}, {Spinrad}, \&
  {Dey}}]{dawson01}
{Dawson}, S., {Stern}, D., {Bunker}, A.~J., {Spinrad}, H., \& {Dey}, A. 2001,
  \aj, 122, 598

\bibitem[{{Faber} {et~al.}(2003)}]{faber03}
{Faber}, S.~M. {et~al.} 2003, in Presented at the Society of Photo-Optical
  Instrumentation Engineers (SPIE) Conference, Vol. 4841, Society of
  Photo-Optical Instrumentation Engineers (SPIE) Conference Series, ed. {M.~Iye
  \& A.~F.~M.~Moorwood}, 1657--1669

\bibitem[{{Frayer} {et~al.}(2006)}]{frayer06}
{Frayer}, D.~T. {et~al.} 2006, \apjl, 647, L9

\bibitem[{{Giavalisco} {et~al.}(2004)}]{giavalisco04}
{Giavalisco}, M. {et~al.} 2004, \apjl, 600, L93

\bibitem[{{Juneau} {et~al.}(2010)}]{juneau10}
{Juneau}, S. {et~al.} 2010, submitted

\bibitem[{{Kobulnicky} \& {Kewley}(2004)}]{kk04}
{Kobulnicky}, H.~A. \& {Kewley}, L.~J. 2004, \apj, 617, 240

\bibitem[{{Laird} {et~al.}(2009)}]{laird09}
{Laird}, E.~S. {et~al.} 2009, \apjs, 180, 102

\bibitem[{{Lowenthal} {et~al.}(1997)}]{lowenthal97}
{Lowenthal}, J.~D. {et~al.} 1997, \apj, 481, 673

\bibitem[{{Magnelli} {et~al.}(2009){Magnelli}, {Elbaz}, {Chary}, {Dickinson},
  {Le Borgne}, {Frayer}, \& {Willmer}}]{magnelli09}
{Magnelli}, B., {Elbaz}, D., {Chary}, R.~R., {Dickinson}, M., {Le Borgne}, D.,
  {Frayer}, D.~T., \& {Willmer}, C.~N.~A. 2009, \aap, 496, 57

\bibitem[{{Magnelli} {et~al.}(2011){Magnelli}, {Elbaz}, {Chary}, {Dickinson},
  {Le Borgne}, {Frayer}, \& {Willmer}}]{magnelli11}
---. 2011, arXiv:1101.2467 [astro-ph]

\bibitem[{{Morrison} {et~al.}(2010){Morrison}, {Owen}, {Dickinson}, {Ivison},
  \& {Ibar}}]{morrison10}
{Morrison}, G.~E., {Owen}, F.~N., {Dickinson}, M., {Ivison}, R.~J., \& {Ibar},
  E. 2010, \apjs, 188, 178

\bibitem[{{Phillips} {et~al.}(1997){Phillips}, {Guzman}, {Gallego}, {Koo},
  {Lowenthal}, {Vogt}, {Faber}, \& {Illingworth}}]{phillips97}
{Phillips}, A.~C., {Guzman}, R., {Gallego}, J., {Koo}, D.~C., {Lowenthal},
  J.~D., {Vogt}, N.~P., {Faber}, S.~M., \& {Illingworth}, G.~D. 1997, \apj,
  489, 543

\bibitem[{{Reddy} {et~al.}(2006){Reddy}, {Steidel}, {Erb}, {Shapley}, \&
  {Pettini}}]{reddy06}
{Reddy}, N.~A., {Steidel}, C.~C., {Erb}, D.~K., {Shapley}, A.~E., \& {Pettini},
  M. 2006, \apj, 653, 1004

\bibitem[{{Riess} {et~al.}(2007)}]{riess07}
{Riess}, A.~G. {et~al.} 2007, \apj, 659, 98

\bibitem[{{Treu} {et~al.}(2005)}]{treu05}
{Treu}, T. {et~al.} 2005, \apj, 633, 174

\bibitem[{{Weiner} {et~al.}(2006)}]{weiner06}
{Weiner}, B.~J. {et~al.} 2006, \apj, 653, 1049

\bibitem[{{Wirth} {et~al.}(2004)}]{wirth04}
{Wirth}, G.~D. {et~al.} 2004, \aj, 127, 3121

\end{thebibliography}
\end{document}